\documentclass[twocolumn,showpacs,aps,prl,nofootinbib]{revtex4}

\usepackage{bm}
\usepackage{epsfig}
\usepackage{amssymb}
\usepackage{amsmath}
\usepackage{slashbox}
\usepackage{calrsfs}
\usepackage{multirow}
\usepackage{rotating}

\begin{document}

\def\be{\begin{equation}}
\def\ee{\end{equation}}

\preprint{DESY~12--???\hspace{13.95cm} ISSN ????--????}
\preprint{September 2012\hspace{16.55cm}}

\title{Gluon- and Quark-Jet Multiplicities with NNNLO and NNLL Accuracy}
\author{P.\ Bolzoni$^{a}$, B.\ A.\ Kniehl$^{a}$, A.V. \ Kotikov$^{a,b}$}
%\affiliation{{II.} Institut f\"ur Theoretische Physik, Universit\"at Hamburg,\\
%             Luruper Chaussee 149, 22761 Hamburg, Germany}
%\author{B.\ A.\ Kniehl}
%\affiliation{{II.} Institut f\"ur Theoretische Physik, Universit\"at Hamburg,\\
%             Luruper Chaussee 149, 22761 Hamburg, Germany}
%\author{A.V. \ Kotikov}
\affiliation{$^{a}${II.} Institut f\"ur Theoretische Physik, Universit\"at Hamburg,\\
             Luruper Chaussee 149, 22761 Hamburg, Germany\\
	     $^{b}$Bogoliubov Laboratory of Theoretical Physics, Joint Institute for Nuclear Research, 
	     141980 Dubna, Russia}
%\affiliation{Bogoliubov Laboratory of Theoretical Physics,
%Joint Institute for Nuclear Research, 141980 Dubna, Russia}

\begin{abstract}
We present a new approach to consider and include both the
perturbative and the non-perturbative contributions to the 
multiplicities of gluon and quark jets. Thanks to this new method, we have 
included for the first time new contributions to these quantities obtaining 
next-to-next-to-leading-logarithmic resummed formulas.
Our analytic expressions depend on two non-perturbative parameters with a 
clear and simple physical interpretation.
A global fit of these two quantities shows how our results 
solve a longstanding discrepancy in the theoretical description of
the data.
%represent an important
%improvement in the explanation of the experimental data.
%describe well
%the data.  
\end{abstract}

\pacs{12.38.Cy,12.39.St,13.66.Bc,13.87.Fh}

\maketitle

%\section{Introduction}
Collisions of particles and nuclei at high energies usually produce
many hadrons.  
In quantum chromodynamics (QCD) their production is due to the
interactions of quarks and gluons and to test it as a theory of
strong interactions, the transition from a description based in terms of quarks and
gluons to the hadrons observed in experiments is always needed.
The production  of hadrons is a typical process where  
non-perturbative phenomena are
involved. However, the hypothesis of local parton-hadron duality  
assumes that parton distributions are simply renormalized in the hadronization
process without changing their shape \cite{Azimov:1984np}, allowing 
perturbative QCD to make predictions. The simplest observables of this kind
are gluon and quark multiplicities
$\langle n_h\rangle_{g}$ and $\langle n_h\rangle_{s}$ which represent
the number of hadrons produced in a gluon and a quark jet respectively. 
In the framework of the generating-functional approach in the modified 
leading logarithmic approximation \cite{Dokshitzer:1991wu}, several
studies of the multiplicities 
have been performed \cite{Malaza:1985jd,Catani:1991pm,Lupia:1997in}.  
In such studies, the ratio  $r=\langle n_h\rangle_g/\langle n_h\rangle_s$
is at least 10\% higher than the data or it has a slope too small. Good agreement 
with the data has been achieved in Ref.~\cite{Eden:1998ig} where recoil effects are
included. Nevertheless in Ref.~\cite{Eden:1998ig} a constant offset to be fitted
to the quark and gluon multiplicities has been introduced, while 
the authors of Ref.~\cite{Capella:1999ms} suggested that
other, better motivated possibilities should be studied.

In this Letter, we study such a possibility inspired by the
new formalism that has recently been proposed in Ref.~\cite{Bolzoni:2012ed}.
Thanks to very recent new results in small-$x$ timelike resummation obtained in Ref.~\cite{Kom:2012hd}, 
we are able to reach the next-to-next-to-leading-logarithmic (NNLL) accuracy level. 
A purely perturbative and analytic prediction has been already attempted
in Ref.~\cite{Capella:1999ms} up to the third order in the expansion parameter 
$\sqrt{\alpha_s}$ \,\emph{i.e.} $\alpha_s^{3/2}$, where paradoxically the quark multiplicity and the
ratio are not well described even if the behavior 
of the perturbative expansion is very good. Our new resummed results that we present here 
are a generalization of what was obtained in Ref.~\cite{Capella:1999ms} and represent
also a solution to this apparent paradox. 

%\section{Formalism}

We consider the standard Mellin-space moments of the coupled gluon-singlet system 
whose evolution in the scale $\mu^2$ is governed in QCD by the Dokshitzer-Gribov-Lipatov-Altarelli-Parisi equations:
\begin{equation}
\label{ap}
\mu^2\frac{d}{d\mu^2} \left(\begin{array}{l} D_s \\ D_g
\end{array}\right)=\left(\begin{array}{ll} P_{qq} & P_{gq} 
\\ P_{qg} & P_{gg}\end{array}\right)\left(\begin{array}{l} D_s \\ D_g
\end{array}\right).
\end{equation}
The timelike splitting functions $P_{ij}$ can be computed
perturbatively in the strong coupling constant:
\begin{equation}
\label{exp}
 P_{ij}(\omega,a_s)=
\sum_{k=0}^\infty a_s^{k+1} P_{ij}^{(k)}(\omega),
\quad  a_s= \frac{\alpha_s}{4\pi},\quad i,j=g,q,
\end{equation}
where $\omega=N-1$ with $N$ being the usual Mellin conjugate variable to the
fraction of longitudinal momentum $x$.
The functions $P_{ij}^{(k)}(\omega)$ with $k=0,1,2$ appearing in Eq.(\ref{exp}) 
in the $\overline{\rm MS}$ scheme can be found in 
Refs.~\cite{Gluck:1992zx,Moch:2007tx,Almasy:2011eq} through next-to-next-to-leading order (NNLO) and in 
Refs.~\cite{Vogt:2011jv,Albino:2011cm,Kom:2012hd} through the NNLL. 

In general it is not possible to diagonalize Eq.(\ref{ap}) because 
the contributions to the splitting function matrix do not commute at 
different orders.
It is, therefore, convenient (see \emph{e.g.}\ Ref.~\cite{Buras:1979yt}) to introduce
a new basis, called plus-minus basis where the LO splitting matrix is diagonal with eigenvalues 
$P_{++}^{(0)}$ and $P_{--}^{(0)}$. We define such a change of basis 
according to the following transformation of the gluon and the singlet fragmentation
functions in Mellin space:
\begin{eqnarray}
D^+(\omega,\mu_0^2) &=& (1-\alpha_{\omega})D_s(\omega,\mu_0^2) 
- \epsilon_\omega D_g(\omega,\mu_0^2),\nonumber\\
D^-(\omega,\mu_0^2))&=& \alpha_{\omega}D_s(\omega,\mu_0^2) + 
\epsilon_\omega D_g(\omega,\mu_0^2)\, , \label{changebasisin}
\end{eqnarray}
where
\begin{equation}
\alpha_\omega=\frac{P_{qq}^{(0)}(\omega)-P_{++}^{(0)}(\omega)}{P_{--}^{(0)}(\omega)-P_{++}^{(0)}(\omega)},
\quad \epsilon_\omega=\frac{P_{gq}^{(0)}(\omega)}{P_{--}^{(0)}(\omega)-P_{++}^{(0)}(\omega)}.
\end{equation}
The general solution to Eq.(\ref{ap}) can be formally written as
\begin{equation}\label{gensol}
D(\mu^2)=T_{\mu^2}\left\{\exp{\int_{\mu_0^2}^{\mu^2}\frac{d\bar{\mu}^2}{\bar{\mu}^2}P(\bar{\mu}^2)}
\right\}D(\mu_0^2),
\end{equation}
where $T_{\mu^2}$ denotes the path ordering with respect to $\mu^2$ and 
\begin{equation}\label{dpm}
D=\left(\begin{array}{l} D^+ \\ D^-
\end{array}\right).
\end{equation}
Now making the following ansatz: 
\begin{eqnarray}\label{ansatz}
&&T_{\mu^2}\left\{\exp{\int_{\mu_0^2}^{\mu^2}\frac{d\bar{\mu}^2}{\bar{\mu}^2}P(\bar{\mu}^2)}
\right\}=\nonumber\\
&&=Z^{-1}(\mu^2)\exp\left[\int_{\mu_0^2}^{\mu^2}\frac{d\bar{\mu}^2}{\bar{\mu}^2}P^{D}
(\bar{\mu}^2)\right]Z(\mu_0^2),
\end{eqnarray}
where
\begin{equation}\label{diagpart}
P^D(\omega)=
\left(\begin{array}{ll} P_{++}(\omega) & 0 
\\ 0 & P_{--}(\omega)\end{array}\right)
\end{equation}
is the all-order diagonal part of the splitting matrix in the plus-minus basis 
and $Z$ is a matrix in the same basis with a
perturbative expansion of the form:
\begin{equation}\label{zpertexp}
Z(\mu^2)=1+a_s(\mu^2)Z^{(1)}+\emph{O}(a_s^2),
\end{equation}
we obtain that
\begin{equation}
\label{decomp}
D_a(\omega,\mu^2)\equiv D_a^+(\omega,\mu^2)+D_a^-(\omega,\mu^2); \qquad a=s,g,
\end{equation} 
where $D_a^+(\omega,\mu^2)$ evolves like a ``plus'' component, $D_a^-(\omega,\mu^2)$
evolves like a ``minus'' component, and 
\begin{equation}
\label{evolsol}
D_a^\pm(\omega,\mu^2)=\tilde{D}_a^\pm(\omega,\mu_0^2)
%\left[\frac{\alpha_s(\mu^2)}{\alpha_s(\mu_0^2)}\right]
%^{-\frac{P_{\pm\pm}^{(0)}}{2\beta_0}}
\hat{T}_{\pm}(\omega,\mu^2,\mu_0^2)
\, H_{a}^\pm(\omega,\mu^2).
\end{equation}
Here $\hat{T}_\pm(\omega,\mu^2,\mu_0^2)$ is a renoramlization group exponent 
which is given by
\begin{equation}\label{rengroupexp}
\hat{T}_{\pm}(\omega,\mu^2,\mu_0^2)  
= \exp \left[\int^{a_s(\mu^2)}_{a_s(\mu_0^2)}
\frac{d\bar{a}_s}{\beta(\bar{a}_s)} \, 
P_{\pm\pm}(\omega,\bar{a}_s) \right] \, ,
\end{equation}
with
\begin{eqnarray}
\beta(a_s(\mu^2))&\equiv&\mu^2\frac{\partial}{\partial\mu^2}a_s(\mu^2)\nonumber\\
&=&-\beta_0 a_s^2(\mu^2)
-\beta_1 a_s^3(\mu^2)+\emph{O}(a_s^4).\label{running}
\end{eqnarray}
We recall that
\begin{eqnarray}
\beta_0 &=& \frac{11}{3}C_A - 
\frac{4}{3}n_f T_R,~~ \nonumber\\
\beta_1 &=& \frac{34}{3}C_A^2 - \frac{20}{3}C_A n_f T_R - 
4 C_F n_f T_R,
\label{1.3}
\end{eqnarray}
where $C_A=3$, $C_F=4/3$ and $T_R=1/2$ in QCD and $n_f$ is
the number of active flavors.
In Eq.(\ref{evolsol}) $H_a^\pm(\omega,\mu^2)$ are perturbative functions 
%\begin{eqnarray}
%\begin{equation}
%\label{pertfun}
%H_a^\pm(\omega,\mu^2) =1- a_s(\mu^2)
%Z_{\pm\mp,a}^{(1)}(\omega)+\emph{O}(a_s^2) \, ,
%\end{equation} 
containing off-diagonal terms of $P$ beyond the LO
and the normalization factors $\tilde{D}_a^\pm(\omega,\mu_0^2)$ 
satisfy the following conditions:
\begin{eqnarray}
\tilde{D}_g^+(\omega,\mu_0^2)&=&-\frac{\alpha_\omega}{\epsilon_\omega}\,
\tilde{D}_s^+(\omega,\mu_0^2)\,;
\nonumber\\
\tilde{D}_g^-(\omega,\mu_0^2)&=&\frac{1-\alpha_\omega}{\epsilon_\omega}\,
\tilde{D}_s^-(\omega,\mu_0^2).
\label{rlo}
\end{eqnarray}
We note that $\tilde{D}^{\pm}_a$ differ from $D^{\pm}_a$ starting at higher orders 
\cite{Buras:1979yt}.
In the following, we collect the resummed formulas. Details of the calculation
will be presented elsewhere \cite{Bolzoni:future}. 

After the resummation is perfomed for $P_{\pm\pm}$ in Eq.(\ref{diagpart}) 
thanks to the results obtained in 
Refs.~\cite{Mueller:1981ex,Gaffney:1984yd,Kom:2012hd},
we find for the first Mellin moment ($\omega=0$) at NNLL:
\begin{eqnarray}
P_{++}^{NNLL}(\omega=0)&=&\gamma_0(1 - K_1 \gamma_0 + K_2  \gamma_0^2+ \emph{O}(\gamma_0^3)),\nonumber\\
P_{--}^{NNLL}(\omega=0)&=&-\frac{8n_f T_R C_F}{3 C_A}\,a_s + \emph{O}(a_s^2),
\end{eqnarray}
where 
\begin{equation}
\gamma_0\equiv P_{++}^{LL}(\omega=0)=\sqrt{2 a_s C_A},
\label{llgamma0}
\end{equation}
and 
\begin{eqnarray}\label{nllfirstA}
K_1&=&  \frac{1}{12} 
\left(11 +4\,\frac{n_f T_R}{C_A} \left(1-\frac{2C_F}{C_A}\right)\right),
\nonumber \\
K_2&=&  \frac{1}{288} 
\left(1193-576\zeta_2 -56\,\frac{n_f T_R}{C_A} 
\left(5+2\frac{C_F}{C_A}\right)\right)\nonumber\\
&& + 16 \frac{n^2_f T^2_R}{C^2_A}
\left(1+4\frac{C_F}{C_A}-12\frac{C^2_F}{C^2_A}\right),
\end{eqnarray}
with $\zeta_2=\pi^2/6$.
Now we can perform the integration in Eq.(\ref{rengroupexp}) up to the NNLL
to obtain that 
\begin{eqnarray} \label{nnllresult}
\hat{T}_{\pm}^{NNLL}(0,Q^2,Q^2_0) = 
\frac{T_{\pm}^{NNLL}(Q^2)}{T_{\pm}^{NNLL}(Q^2_0)},\,\,\,\,\,\,\,\,\,\,\,\,\,\,\,\,\,&&\\ 
T_{+}^{NNLL}(Q^2) = \exp\Bigl\{\frac{4C_A}{\beta_0 \gamma^0(Q^2)}
\Bigl[1+ \nonumber \,\,\,\,\,\,\,\,\,\,\,\,\,\,\,\,\,&&\\
+\Bigl(b_1-2C_AK_2\Bigr)a_s(Q^2)\Bigr]\Bigr\}
{\Bigl(a_s(Q^2)\Bigr)}^{d_+},&&\\
T_{-}^{NNLL}(Q^2)={\Bigl(a_s(Q^2)\Bigr)}^{d_-},\,\,\,\,\,\,\,\,\,\,\,\,\,\,\,\,\,\,\,\,\,\,\,\,\,\,\,\,\,\,\,\,\,\,&&
\end{eqnarray}
where
\begin{equation} b_1=\beta_1/\beta_0,~~
d_{-} = \frac{8 n_f T_R C_F}{3 C_A \beta_0},~~ d_{+} = \frac{2 C_A K_1}{\beta_0}. 
\label{anomdim}
\end{equation}

We are now ready to define the avarage multiplicities in our formalism: 
\begin{equation}\label{multdef}
\langle n_h(Q^2)\rangle_a\equiv D_a(0,Q^2)= D_a^+(0,Q^2) + D_a^-(0,Q^2),
\end{equation}
with $a=g,s$ for the gluon and quark multiplicities, respectively.
From Eqs.(\ref{evolsol}) and (\ref{rlo}) we have that
\begin{equation}
\label{evolsola}
\frac{D_g^+(0,Q^2)}{D_s^+(0,Q^2)}=-\lim_{\omega\rightarrow 0}\frac{\alpha_\omega}{\epsilon_\omega}
\frac{H^+_g(\omega,Q^2)}{H^+_s(\omega,Q^2)}\equiv r_+(Q^2),
\end{equation}
and
\begin{equation}
\label{rmin}
\frac{D_g^-(0,Q^2)}{D_s^-(0,Q^2)}=\lim_{\omega\rightarrow 0}
\frac{1-\alpha_\omega}{\epsilon_\omega}
\frac{H^-_g(\omega,Q^2)}{H^-_s(\omega,Q^2)}\equiv r_-(Q^2).
\end{equation}
Using these definitions, it is convenient  to write for the gluon and
quark multiplicities in general:
\begin{eqnarray}
\langle n_h(Q^2)\rangle_g=\tilde{D}_g^+(0,Q_0^2)\hat{T}_+^{res}(0,Q^2,Q_0^2)
H^+_g(0,Q^2)&&\nonumber\\
\qquad\qquad+\tilde{D}_s^-(0,Q_0^2)r_-(Q^2)\hat{T}_-^{res}(0,Q^2,Q_0^2)
H^-_s(0,Q^2),&&\nonumber\\
\langle n_h(Q^2)\rangle_s=\frac{\tilde{D}_g^+(0,Q_0^2)}{r_+(Q^2)}\hat{T}_+^{res}(0,Q^2,Q_0^2)
H^+_g(0,Q^2)&&\nonumber\\
+\tilde{D}_s^-(0,Q_0^2)\hat{T}_-^{res}(0,Q^2,Q_0^2)H^-_s(0,Q^2).\,\,\,\,\,&&
\label{multgen}
\end{eqnarray}
For the coefficients of the renormalization group
exponents, we clearly have the following simple relations
at the lowest order in $a_s$ 
\begin{eqnarray}
&&r_+(Q^2)=C_A/C_F;\qquad r_-(Q^2)=0; \nonumber\\ &&H^{\pm}_s(0,Q^2)=1;\qquad 
\tilde{D}_a^\pm(0,Q_0^2)=D_a^\pm(0,Q_0^2),\label{lonnll}
\end{eqnarray}
with $a=g,s$.
One would like to include higher-order corrections to Eq.(\ref{lonnll}).
However, this is highly non-trivial because the general perturbative structures
of the functions $H^{\pm}_a$ and $Z_{\pm\mp,a}$, whose knowledge is required for 
the resummation, are not known. Fortunatly, general assumptions and approximations can be made to 
improve them. Firstly, it is a
well known fact that the plus components by themselves represent the dominant
contributions for both the gluon and the quark multiplicities 
(see \emph{e.g.} Refs.~\cite{Schmelling:1994py,Dremin:2000ep}). Secondly, Eq.(\ref{rmin}) tells us that $D^-_g$ is
suppressed with respect to $D^-_s$, because $\alpha_\omega\sim
1+\emph{O}(\omega)$. These two facts suggest us that to keep $r_-(Q^2)=0$ 
even at higher orders should still represent a good approximation.
Then we notice that higher-order corrections to $\tilde{D}^\pm_a(0,Q_0^2)$ and $H^{\pm}_a(0,Q^2)$ 
just represent a redefinition of 
$D^\pm_a(0,Q_0^2)$ apart from running coupling effects starting at order $a_s^2$. 
Therefore 
we assume that these corrections can be neglected.
Now we can finally discuss higher-order corrections to $r_+(Q^2)$, which
represents the ratio of the pure plus components. Accordingly, we can
intepret the result in Eq.(5) of Ref.~\cite{Capella:1999ms} as higher-order 
corrections to Eq.(\ref{evolsola}). This interpretation is explicitly 
confirmed up to order $a_s$ in Chapter 7 of Ref.~\cite{Dokshitzer:1991wu}, where also the
same set of equations used in the computation of Ref.~\cite{Capella:1999ms} are obtained. 
Further arguments to support it and its scheme dependence will be discussed in Ref.~\cite{Bolzoni:future}.
%However a comment is needed because identifying the quantity $r_+(Q^2)$ as
%the one computed in \cite{Capella:1999ms} we are assuming the scheme dependence
%to be very small and this is reasonable because in \cite{Albino:2011cm} we have
%shown the scheme independence up to the NLL.
We denote the approximation in which Eqs.(\ref{nnllresult},\ref{lonnll})
are used as $\rm{LO+NNLL}$ and the one in which $r_+(Q^2)$ in Eq.(\ref{lonnll}) is
replaced by the result of Eq.(5) 
in Ref.~\cite{Capella:1999ms} up to order $a_s^{3/2}$ as 
$\rm{NNNLO}_{\rm{approx}}+\rm{NNLL}$. That this last one is actually a good 
approximation will be shown below.
\begin{figure}
%\centering
%\begin{minipage}[b]{6.9cm}
%\centering
\includegraphics[scale=0.6]{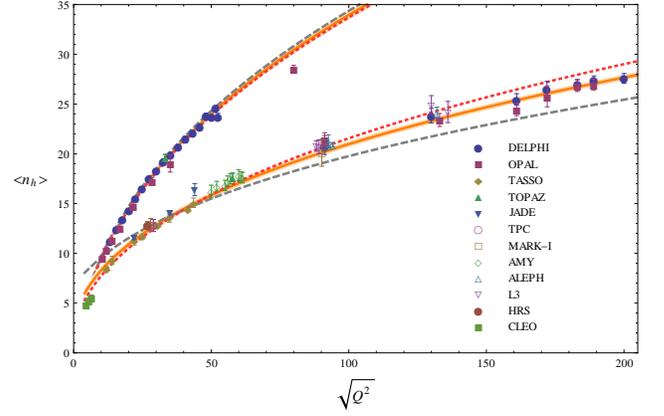}
\caption{\footnotesize{Gluon and quark multiplicities fits compared to the data. The gray dashed line is the
 $\rm{LO}+\rm{NNLL}$ result, the orange solid line is the $\rm{NNNLO}_{\rm{approx}}+\rm{NNLL}$ 
 result and the red dotted line is the fit with four constant coefficients. The orange band
corresponds to the estimated error of the fitted parameters in the
$\rm{NNNLO}_{\rm{approx}}+\rm{NNLL}$ case.}}
\label{Fig:plotmult}
%\end{minipage}
\end{figure}
\begin{figure}
%\centering
%\begin{minipage}[b]{6.9cm}
%\centering
\includegraphics[scale=0.6]{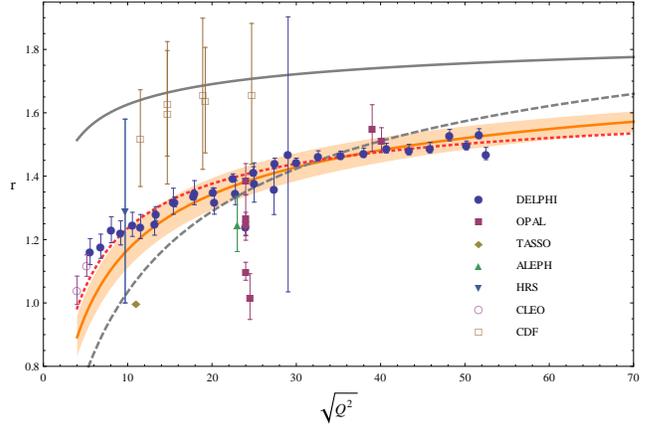}
\caption{\footnotesize{Glun-quark multiplicity ratio prediction compared to data. The 
gray solid upper line is the prediction of Ref.~\cite{Capella:1999ms}, the others are as in Fig.\ref{Fig:plotmult}}}
\label{Fig:ratio}
%\end{minipage}
\end{figure}
In both approximations considered we can summarize the main
theoretical result of this Letter
in the following way
\begin{eqnarray}
\langle
n_h(Q^2)\rangle_g=D_g(0,Q_0^2)\hat{T}_+^{res}(0,Q^2,Q_0^2),\,\,\,\,\,\,\,
\,\,\,\,\,\,&&\label{mainresult1}\\
\langle n_h(Q^2)\rangle_s=D_g(0,Q_0^2)\frac{\hat{T}_+^{res}(0,Q^2,Q_0^2)}
{r_+(Q^2)}\,\,\,\,\,\,\,\,\,\,\,\,\,&&\nonumber\\
+\left[D_s(0,Q_0^2)-\frac{D_g(0,Q_0^2)}{{r_+(Q_0^2)}}\right]\hat{T}_-^{res}(0,Q^2,Q_0^2),&&
\label{mainresult2}
\end{eqnarray}
for the multiplicities, and 
\begin{eqnarray}\label{finalratio}
&&r(Q^2) \equiv\frac{\langle n_h(Q^2)\rangle_g}{\langle n_h(Q^2)\rangle_s} \\
&&= 
\frac{r_{+}(Q^2)}{\left[1 + \frac{r_{+}(Q^2)}{r_{+}(Q_0^2)}\left(
\frac{D_s(0,Q^2_0)r_{+}(Q_0^2)}{D_g(0,Q^2_0)}
-1 \right)
\frac{\hat{T}_{-}^{res}(0,Q^2,Q^2_0)}{\hat{T}_{+}^{res}(0,Q^2,Q^2_0)}\right]} \, ,
\nonumber  
\end{eqnarray}
for the gluon-quark multiplicity ratio. 
Equations (\ref{mainresult1},\ref{mainresult2}) depend
only on two parameters, $D_g(0,Q^2_0)$ and 
$D_s(0,Q^2_0)$, with a simple physical interpretation: they are just the gluon
and the quark multiplicities at the arbitrary scale $Q_0$.

We have performed a global fit of our resummed
formulas, Eqs.(\ref{mainresult1},\ref{mainresult2}), to the experimental data
to extract the values of $D_g(0,Q^2_0)$ and $D_s(0,Q^2_0)$.
With $Q_0=50 \,\rm{GeV}$, the result of the fit   
is given by
\begin{eqnarray}
D_g(0,Q_0^2)&=&24.31\pm0.85; \quad\rm{ 90\%\,\,\, C.L.}\nonumber\\
D_s(0,Q_0^2)&=&15.49\pm0.90; \quad\rm{ 90\%\,\,\, C.L.},
\label{fitlonnll}
\end{eqnarray}
in the $\rm{LO}+\rm{NNLL}$ case
and by
\begin{eqnarray}
D_g(0,Q_0^2)&=&24.02\pm0.36; \quad\rm{ 90\%\,\,\,C.L.}\nonumber\\
D_s(0,Q_0^2)&=&15.83\pm0.37;\quad\rm{ 90\%\,\,\,C.L.},
\label{fitnnll}
\end{eqnarray}
in the $\rm{NNNLO}_{\rm{approx}}+\rm{NNLL}$ case
in agreement with the experimental values within the errors.
However, the $90\%$ C.L. error in the $\rm{NNNLO}_{\rm{approx}}+\rm{NNLL}$
case is much smaller reflecting a much better fit to the data at all energies. Indeed,
per degree of freedom we obtain $\chi^2=18.09$ in the $\rm{LO}+\rm{NNLL}$ case, while we have $\chi^2=3.71$
in the $\rm{NNNLO}_{\rm{approx}}+\rm{NNLL}$ case. In our analysis, we have used the next-to-leading order 
solution for the running coupling according to Eq.(\ref{running}) with $\alpha_s(M_Z)=0.118$ and
$n_f=5$. We have checked that varying the arbitrary scale $Q_0^2$ does not change the resulting 
value of $\chi^2$ as expected and that moving from LL to NNLL
the renormalization scale dependence is strongly reduced.

In Fig.\ref{Fig:plotmult} we plot the gluon and quark multiplicities according to 
Eqs.(\ref{multgen},\ref{mainresult1},\ref{mainresult2}) using the fitted parameters 
given in Eqs.(\ref{fitlonnll},\ref{fitnnll}). Using the data selection of
Ref.~\cite{Abdallah:2005cy}, the measurements are taken from 
Refs.~\cite{Abdallah:2005cy,Abbiendi:1999pi,Nakabayashi:1997hr,Abbiendi:2004pr} for the
gluon multiplicity and from Refs.~\cite{Siebel:2003zz,Alam:1997ht} and references therein for
the quark multiplicity. The result of a fit where the normalization coefficients 
are assumed constant without any additional constraint is also plotted 
showing that $\rm{NNNLO}_{\rm{approx}}+\rm{NNLL}$
is indeed a good approximation.
To check the consistency of the data sets, we have used Eq.(\ref{finalratio}) together with the result of the
fit from the gluon and quark multiplicities in Eqs.(\ref{fitlonnll}) and (\ref{fitnnll}) to predict the gluon-quark
multiplicity ratio. The result together with the corresponding data are shown 
in Fig.\ref{Fig:ratio}.  The data are taken from 
Refs.~\cite{Albrecht:1991vp,Alam:1992ir,Alam:1997ht,Abbiendi:1999pi,Abreu:1999rs,Acosta:2004js,Abdallah:2005cy,Siebel:2003zz}
and references therein, covering essentially all available measurements.
One can see that the data do not agree very well at small scales, an 
isssue that will be discussed elsewhere \cite{Bolzoni:future}.

As concluding remarks we remind here that the main 
problem in describing the data was that 
the theory failed badly in the description of the data for the gluon and the 
quark jets simultaneously (or equivalently for the ratio $r$) even if the perturbative 
series seems to converge very well. We have shown in this Letter 
that our $\rm{NNNLO}_{\rm{approx}}+\rm{NNLL}$ result solves this
problem explaining the discrepancy of the results with the data obtained 
in Ref.~\cite{Capella:1999ms} as due to the absence of the singlet ``minus'' component
governed by $\hat{T}_-^{res}(0,Q^2,Q_0^2)$ in Eqs.(\ref{mainresult2}) and (\ref{finalratio}). 
This component is included here for the first time.
The most natural possible future improvement consists in including corrections
of next-to-leading order or beyond 
to $r_-(Q^2)$.
Our generalized result depends on two parameters, which represent our initial condition. 
They have been fixed performing a fit and have a simple physical meaning because they just
represent the gluon and the quark multiplicity at a certain arbitrary scale $Q_0$.
We hope that additional measurements of these observables will come from the LHC
to test our results on a much wider energy range. 

We kindly thank A.~Vogt and G.~Kramer for valuable discussions.
The work of A.V.K.\ was supported in part by the Heisenberg-Landau program and the Russian Foundation for Basic
Research RFBR through Grant No.\ 11--02--01454--a.
This work was supported in part by the German Federal Ministry for Education
and Research BMBF through Grant No.\ 05H12GUE, by the German Research
Foundation DFG through Grant No. KN 365/5-3, and by the Helmholtz Association HGF through the Helmholtz
Alliance Ha~101 {\it Physics at the Terascale}.


\begin{thebibliography}{30}
\expandafter\ifx\csname natexlab\endcsname\relax\def\natexlab#1{#1}\fi
\expandafter\ifx\csname bibnamefont\endcsname\relax
  \def\bibnamefont#1{#1}\fi
\expandafter\ifx\csname bibfnamefont\endcsname\relax
  \def\bibfnamefont#1{#1}\fi
\expandafter\ifx\csname citenamefont\endcsname\relax
  \def\citenamefont#1{#1}\fi
\expandafter\ifx\csname url\endcsname\relax
  \def\url#1{\texttt{#1}}\fi
\expandafter\ifx\csname urlprefix\endcsname\relax\def\urlprefix{URL }\fi
\providecommand{\bibinfo}[2]{#2}
\providecommand{\eprint}[2][]{\url{#2}}

\bibitem[{\citenamefont{Azimov et~al.}(1985)\citenamefont{Azimov, Dokshitzer,
  Khoze, and Troyan}}]{Azimov:1984np}
\bibinfo{author}{\bibfnamefont{Y.~I.} \bibnamefont{Azimov}},
  \bibinfo{author}{\bibfnamefont{Y.~L.} \bibnamefont{Dokshitzer}},
  \bibinfo{author}{\bibfnamefont{V.~A.} \bibnamefont{Khoze}}, \bibnamefont{and}
  \bibinfo{author}{\bibfnamefont{S.~I.} \bibnamefont{Troyan}},
  \bibinfo{journal}{Z. Phys.} \textbf{\bibinfo{volume}{C27}},
  \bibinfo{pages}{65} (\bibinfo{year}{1985}).

\bibitem[{\citenamefont{Dokshitzer et~al.}(1991)\citenamefont{Dokshitzer,
  Khoze, Mueller, and Troian}}]{Dokshitzer:1991wu}
\bibinfo{author}{\bibfnamefont{Y.~L.} \bibnamefont{Dokshitzer}},
  \bibinfo{author}{\bibfnamefont{V.~A.} \bibnamefont{Khoze}},
  \bibinfo{author}{\bibfnamefont{A.~H.} \bibnamefont{Mueller}},
  \bibnamefont{and} \bibinfo{author}{\bibfnamefont{S.~I.} \bibnamefont{Troian}}
  , \bibinfo{note}{\emph{Basics of Perturbative QCD}(Frontiers, Gif-sur-Yvette, France, 1991),
  p.274}.

\bibitem[{\citenamefont{Malaza and Webber}(1986)}]{Malaza:1985jd}
\bibinfo{author}{\bibfnamefont{E.~D.} \bibnamefont{Malaza}} \bibnamefont{and}
  \bibinfo{author}{\bibfnamefont{B.~R.} \bibnamefont{Webber}},
  \bibinfo{journal}{Nucl. Phys.} \textbf{\bibinfo{volume}{B267}},
  \bibinfo{pages}{702} (\bibinfo{year}{1986}).

\bibitem[{\citenamefont{Catani et~al.}(1992)\citenamefont{Catani, Dokshitzer,
  Fiorani, and Webber}}]{Catani:1991pm}
\bibinfo{author}{\bibfnamefont{S.}~\bibnamefont{Catani}},
  \bibinfo{author}{\bibfnamefont{Y.~L.} \bibnamefont{Dokshitzer}},
  \bibinfo{author}{\bibfnamefont{F.}~\bibnamefont{Fiorani}}, \bibnamefont{and}
  \bibinfo{author}{\bibfnamefont{B.~R.} \bibnamefont{Webber}},
  \bibinfo{journal}{Nucl. Phys.} \textbf{\bibinfo{volume}{B377}},
  \bibinfo{pages}{445} (\bibinfo{year}{1992}).

\bibitem[{\citenamefont{Lupia and Ochs}(1998)}]{Lupia:1997in}
\bibinfo{author}{\bibfnamefont{S.}~\bibnamefont{Lupia}} \bibnamefont{and}
  \bibinfo{author}{\bibfnamefont{W.}~\bibnamefont{Ochs}},
  \bibinfo{journal}{Phys. Lett.} \textbf{\bibinfo{volume}{B418}},
  \bibinfo{pages}{214} (\bibinfo{year}{1998}), \eprint{hep-ph/9707393}.

\bibitem[{\citenamefont{Eden and Gustafson}(1998)}]{Eden:1998ig}
\bibinfo{author}{\bibfnamefont{P.}~\bibnamefont{Eden}} \bibnamefont{and}
  \bibinfo{author}{\bibfnamefont{G.}~\bibnamefont{Gustafson}},
  \bibinfo{journal}{JHEP} \textbf{\bibinfo{volume}{09}}, \bibinfo{pages}{015}
  (\bibinfo{year}{1998}), \eprint{hep-ph/9805228}.

\bibitem[{\citenamefont{Capella et~al.}(2000)\citenamefont{Capella, Dremin,
  Gary, Nechitailo, and Tran Thanh~Van}}]{Capella:1999ms}
\bibinfo{author}{\bibfnamefont{A.}~\bibnamefont{Capella}},
  \bibinfo{author}{\bibfnamefont{I.~M.} \bibnamefont{Dremin}},
  \bibinfo{author}{\bibfnamefont{J.~W.} \bibnamefont{Gary}},
  \bibinfo{author}{\bibfnamefont{V.~A.} \bibnamefont{Nechitailo}},
  \bibnamefont{and} \bibinfo{author}{\bibfnamefont{J.}~\bibnamefont{Tran
  Thanh~Van}}, \bibinfo{journal}{Phys. Rev.} \textbf{\bibinfo{volume}{D61}},
  \bibinfo{pages}{074009} (\bibinfo{year}{2000}), \eprint{hep-ph/9910226}.

\bibitem[{\citenamefont{Bolzoni}(2012)}]{Bolzoni:2012ed}
\bibinfo{author}{\bibfnamefont{P.}~\bibnamefont{Bolzoni}}
  (\bibinfo{year}{2012}), \eprint{1206.3039}.

\bibitem[{\citenamefont{Kom et~al.}(2012)\citenamefont{Kom, Vogt, and
  Yeats}}]{Kom:2012hd}
\bibinfo{author}{\bibfnamefont{C.-H.} \bibnamefont{Kom}},
  \bibinfo{author}{\bibfnamefont{A.}~\bibnamefont{Vogt}}, \bibnamefont{and}
  \bibinfo{author}{\bibfnamefont{K.}~\bibnamefont{Yeats}}
  (\bibinfo{year}{2012}), \eprint{1207.5631}.

\bibitem[{\citenamefont{Gluck et~al.}(1993)\citenamefont{Gluck, Reya, and
  Vogt}}]{Gluck:1992zx}
\bibinfo{author}{\bibfnamefont{M.}~\bibnamefont{Gluck}},
  \bibinfo{author}{\bibfnamefont{E.}~\bibnamefont{Reya}}, \bibnamefont{and}
  \bibinfo{author}{\bibfnamefont{A.}~\bibnamefont{Vogt}},
  \bibinfo{journal}{Phys. Rev.} \textbf{\bibinfo{volume}{D48}},
  \bibinfo{pages}{116} (\bibinfo{year}{1993}).

\bibitem[{\citenamefont{Moch and Vogt}(2008)}]{Moch:2007tx}
\bibinfo{author}{\bibfnamefont{S.}~\bibnamefont{Moch}} \bibnamefont{and}
  \bibinfo{author}{\bibfnamefont{A.}~\bibnamefont{Vogt}},
  \bibinfo{journal}{Phys. Lett.} \textbf{\bibinfo{volume}{B659}},
  \bibinfo{pages}{290} (\bibinfo{year}{2008}), \eprint{0709.3899}.

\bibitem[{\citenamefont{Almasy et~al.}(2011)\citenamefont{Almasy, Vogt, and
  Moch}}]{Almasy:2011eq}
\bibinfo{author}{\bibfnamefont{A.~A.} \bibnamefont{Almasy}},
  \bibinfo{author}{\bibfnamefont{A.}~\bibnamefont{Vogt}}, \bibnamefont{and}
  \bibinfo{author}{\bibfnamefont{S.}~\bibnamefont{Moch}}
  (\bibinfo{year}{2011}), \eprint{1107.2263}.

\bibitem[{\citenamefont{Vogt}(2011)}]{Vogt:2011jv}
\bibinfo{author}{\bibfnamefont{A.}~\bibnamefont{Vogt}}, \bibinfo{journal}{JHEP}
  \textbf{\bibinfo{volume}{10}}, \bibinfo{pages}{025} (\bibinfo{year}{2011}),
  \eprint{1108.2993}.

\bibitem[{\citenamefont{Albino et~al.}(2012)\citenamefont{Albino, Bolzoni,
  Kniehl, and Kotikov}}]{Albino:2011cm}
\bibinfo{author}{\bibfnamefont{S.}~\bibnamefont{Albino}},
  \bibinfo{author}{\bibfnamefont{P.}~\bibnamefont{Bolzoni}},
  \bibinfo{author}{\bibfnamefont{B.~A.} \bibnamefont{Kniehl}},
  \bibnamefont{and} \bibinfo{author}{\bibfnamefont{A.~V.}
  \bibnamefont{Kotikov}}, \bibinfo{journal}{Nucl. Phys.}
  \textbf{\bibinfo{volume}{B855}}, \bibinfo{pages}{801} (\bibinfo{year}{2012}),
  \eprint{1108.3948}.

\bibitem[{\citenamefont{Buras}(1980)}]{Buras:1979yt}
\bibinfo{author}{\bibfnamefont{A.~J.} \bibnamefont{Buras}},
  \bibinfo{journal}{Rev. Mod. Phys.} \textbf{\bibinfo{volume}{52}},
  \bibinfo{pages}{199} (\bibinfo{year}{1980}).

\bibitem[{\citenamefont{Bolzoni et~al.}({to be published})\citenamefont{Bolzoni,
  Kniehl, and Kotikov}}]{Bolzoni:future}
\bibinfo{author}{\bibfnamefont{P.}~\bibnamefont{Bolzoni}},
  \bibinfo{author}{\bibfnamefont{B.~A.} \bibnamefont{Kniehl}},
  \bibnamefont{and} \bibinfo{author}{\bibfnamefont{A.}~\bibnamefont{Kotikov}}
  (\bibinfo{year}{{to be published}}).

\bibitem[{\citenamefont{Mueller}(1981)}]{Mueller:1981ex}
\bibinfo{author}{\bibfnamefont{A.~H.} \bibnamefont{Mueller}},
  \bibinfo{journal}{Phys. Lett.} \textbf{\bibinfo{volume}{B104}},
  \bibinfo{pages}{161} (\bibinfo{year}{1981}).

\bibitem[{\citenamefont{Gaffney and Mueller}(1985)}]{Gaffney:1984yd}
\bibinfo{author}{\bibfnamefont{J.~B.} \bibnamefont{Gaffney}} \bibnamefont{and}
  \bibinfo{author}{\bibfnamefont{A.~H.} \bibnamefont{Mueller}},
  \bibinfo{journal}{Nucl. Phys.} \textbf{\bibinfo{volume}{B250}},
  \bibinfo{pages}{109} (\bibinfo{year}{1985}).

\bibitem[{\citenamefont{Schmelling}(1995)}]{Schmelling:1994py}
\bibinfo{author}{\bibfnamefont{M.}~\bibnamefont{Schmelling}},
  \bibinfo{journal}{Phys.Scripta} \textbf{\bibinfo{volume}{51}},
  \bibinfo{pages}{683} (\bibinfo{year}{1995}).

\bibitem[{\citenamefont{Dremin and Gary}(2001)}]{Dremin:2000ep}
\bibinfo{author}{\bibfnamefont{I.~M.} \bibnamefont{Dremin}} \bibnamefont{and}
  \bibinfo{author}{\bibfnamefont{J.~W.} \bibnamefont{Gary}},
  \bibinfo{journal}{Phys. Rept.} \textbf{\bibinfo{volume}{349}},
  \bibinfo{pages}{301} (\bibinfo{year}{2001}), \eprint{hep-ph/0004215}.

\bibitem[{\citenamefont{Abdallah et~al.}(2005)}]{Abdallah:2005cy}
\bibinfo{author}{\bibfnamefont{J.}~\bibnamefont{Abdallah}} \bibnamefont{et~al.}
  (\bibinfo{collaboration}{DELPHI Collaboration}),
  \bibinfo{journal}{Eur.Phys.J.} \textbf{\bibinfo{volume}{C44}},
  \bibinfo{pages}{311} (\bibinfo{year}{2005}), \eprint{hep-ex/0510025}.

\bibitem[{\citenamefont{Abbiendi et~al.}(1999)}]{Abbiendi:1999pi}
\bibinfo{author}{\bibfnamefont{G.}~\bibnamefont{Abbiendi}} \bibnamefont{et~al.}
  (\bibinfo{collaboration}{OPAL Collaboration}), \bibinfo{journal}{Eur. Phys. J.}
  \textbf{\bibinfo{volume}{C11}}, \bibinfo{pages}{217} (\bibinfo{year}{1999}),
  \eprint{hep-ex/9903027}.

\bibitem[{\citenamefont{Nakabayashi et~al.}(1997)}]{Nakabayashi:1997hr}
\bibinfo{author}{\bibfnamefont{K.}~\bibnamefont{Nakabayashi}}
  \bibnamefont{et~al.} (\bibinfo{collaboration}{TOPAZ Collaboration}),
  \bibinfo{journal}{Phys.Lett.} \textbf{\bibinfo{volume}{B413}},
  \bibinfo{pages}{447} (\bibinfo{year}{1997}).

\bibitem[{\citenamefont{Abbiendi et~al.}(2004)}]{Abbiendi:2004pr}
\bibinfo{author}{\bibfnamefont{G.}~\bibnamefont{Abbiendi}} \bibnamefont{et~al.}
  (\bibinfo{collaboration}{OPAL Collaboration}), \bibinfo{journal}{Eur.Phys.J.}
  \textbf{\bibinfo{volume}{C37}}, \bibinfo{pages}{25} (\bibinfo{year}{2004}),
  \eprint{hep-ex/0404026}.

\bibitem[{\citenamefont{Siebel}(2003)}]{Siebel:2003zz}
\bibinfo{author}{\bibfnamefont{M.}~\bibnamefont{Siebel}}
  (\bibinfo{year}{2003}), \bibinfo{note}{PHD thesis,Bergische Universita¨t Wuppertal [Report No. wU-B-DIS-2003-11],
   http://elpub.bib
.uni-wuppertal.de/servlets/DerivateServlet/Derivate-973/
dc0301.pdf.}.

\bibitem[{\citenamefont{Alam et~al.}(1997)}]{Alam:1997ht}
\bibinfo{author}{\bibfnamefont{M.~S.} \bibnamefont{Alam}} \bibnamefont{et~al.}
  (\bibinfo{collaboration}{CLEO Collaboration}), \bibinfo{journal}{Phys. Rev.}
  \textbf{\bibinfo{volume}{D56}}, \bibinfo{pages}{17} (\bibinfo{year}{1997}),
  \eprint{hep-ex/9701006}.

\bibitem[{\citenamefont{Albrecht et~al.}(1992)}]{Albrecht:1991vp}
\bibinfo{author}{\bibfnamefont{H.}~\bibnamefont{Albrecht}} \bibnamefont{et~al.}
  (\bibinfo{collaboration}{ARGUS Collaboration}), \bibinfo{journal}{Z. Phys.}
  \textbf{\bibinfo{volume}{C54}}, \bibinfo{pages}{13} (\bibinfo{year}{1992}).

\bibitem[{\citenamefont{Alam et~al.}(1992)}]{Alam:1992ir}
\bibinfo{author}{\bibfnamefont{M.~S.} \bibnamefont{Alam}} \bibnamefont{et~al.}
  (\bibinfo{collaboration}{CLEO Collaboration}), \bibinfo{journal}{Phys. Rev.}
  \textbf{\bibinfo{volume}{D46}}, \bibinfo{pages}{4822} (\bibinfo{year}{1992}).

\bibitem[{\citenamefont{Abreu et~al.}(1999)}]{Abreu:1999rs}
\bibinfo{author}{\bibfnamefont{P.}~\bibnamefont{Abreu}} \bibnamefont{et~al.}
  (\bibinfo{collaboration}{DELPHI Collaboration}),
  \bibinfo{journal}{Phys.Lett.} \textbf{\bibinfo{volume}{B449}},
  \bibinfo{pages}{383} (\bibinfo{year}{1999}), \eprint{hep-ex/9903073}.

\bibitem[{\citenamefont{Acosta et~al.}(2005)}]{Acosta:2004js}
\bibinfo{author}{\bibfnamefont{D.}~\bibnamefont{Acosta}} \bibnamefont{et~al.}
  (\bibinfo{collaboration}{CDF Collaboration}),
  \bibinfo{journal}{Phys.Rev.Lett.} \textbf{\bibinfo{volume}{94}},
  \bibinfo{pages}{171802} (\bibinfo{year}{2005}).

\end{thebibliography}
\end{document}